\documentclass[pre,floatfix,showpacs]{revtex4}

\usepackage{graphics}
\def\be{\begin{equation}}
\def\ee{\end{equation}}
\def\bea{\begin{eqnarray}}
\def\eea{\end{eqnarray}}

\begin{document}
\title{Driven response of time delay coupled limit cycle oscillators}
\author{D. V. Ramana Reddy}
\altaffiliation{Present address: Center for Neural Science, New York University, New York, NY 10003.}
\author{Abhijit Sen}
\author{George L. Johnston}
\altaffiliation{Present address: EduTron Corp., 5 Cox Road, Winchester, MA 01890.}
\affiliation{Institute for Plasma Research, Bhat, Gandhinagar 382428, Gujarat, India.}

\begin{abstract}
We study the periodic forced response of a system of two limit
cycle oscillators that interact with each other via a time
delayed coupling. Detailed bifurcation diagrams in the parameter
space of the forcing amplitude and forcing frequency are obtained
for various interesting limits using numerical and analytical
means. In particular, the effects of the coupling strength, the
natural frequency spread of the two oscillators and the time delay
parameter on the size and nature of the entrainment domain are
delineated. For an appropriate choice of time delay, synchronization
can occur with infinitesimal forcing amplitudes even at off-resonant driving.
The system is also found to display a nonlinear response
on certain critical contours in the space of the coupling strength
and time delay. Numerical simulations with a large number of coupled driven 
oscillators display similar behavior. Time delay offers a novel tuning knob for
controlling the system response over a wide range of frequencies
and this may have important practical applications. \\ 

\end{abstract}

\maketitle

\section{Introduction}
Coupled limit cycle oscillators have been extensively
studied in recent times to understand synchronization
phenomena in various physical, chemical and biological
systems \cite{Win:80book,Str:00,ARSBHSL:00,BS:00,Str:01,TTYNFE:01}.
Examples include coupled lasers \cite{WW:88,WBJR:90,TMRCLE:99},
coupled magnetrons \cite{BSWSH:89}, arrays of Josephson 
junctions \cite{LPT:66,CIKL:86,MT:89,BRTL:90,ST:94,CS:95,WCS:96},
coupled chemical 
reactors \cite{Kur:76,Neu:80,NS:80,Kur:81,CE:89,HE:89,Bar:90,YYM:93}, 
and coupled arrays of biological cells \cite{HKK:95,Izh:99,BSher:00}.
Many of these practical systems are also often subject to
external driving forces.
For example, the day and night variation of
the solar radiation input provides a natural
periodic forcing of many biological and ecological systems.
Electronic pace-maker devices implanted in the human body for regulating
cardiac rhythm, driven chemical oscillations in industrial
reactors, and control of chaos through periodic forcing of
mechanical systems are a few other common examples. Another
ubiquitous feature of natural systems is the presence of time
delay in the mutual interaction between their component
elements \cite{TKRGH:96,MY:99}.
Such delays are normally associated with finite propagation times
of signals, finite reaction times in chemical systems, or
individual neuron firing times in neural networks.
Time delay
introduces interesting new features in the collective dynamics
of coupled limit cycle oscillators as has been pointed out in a
number of past studies \cite{SW:89,NSK:91,NTM:94,KPR:97,%
RSJ:98,RSJ:99,FV:97,CW:98,Pyr:98,WeiR:99,YS:99,BC:99,%
CKKH:00,KVM:00,WR:99,SR:00}.\\

The characterization of the driven response of coupled oscillator
systems has important practical applications, and has been carried out in
the past for a number of simple systems \cite{GH:83book,CFW:97}.
A well known case is
the driven Kuramoto model where the collective phase evolution
of a group of weakly coupled limit cycle oscillators due to periodic forcing
has been examined \cite{CKH:94}. However for many practical applications
the approximation of weak coupling is not valid. For such cases where the coupling is strong,
one needs to take into account the amplitude evolution in addition to the
phase dynamics. Examples include the Hodgkin-Huxley family of equations 
used in many biological applications and coupled Stuart-Landau oscillators
which have been investigated in the past in the context of spontaneous collective synchronization
phenomena \cite{AEK:90,MMS:91}. Recently we have generalized the Stuart-Landau model to include
time delay effects in the coupling terms \cite{RSJ:98,RSJ:99}. In this paper we
study the driven periodic response of such a generalized system
when it is subjected to an external oscillating force. For
simplicity we restrict ourselves mostly to the investigation of just two delay coupled
oscillators which are each subjected to an identical periodic
force. However as we briefly demonstrate toward the end of the paper the
results hold for a system of large number of oscillators as well. Our main objective 
is to examine the nature of the synchronized
response of the system in various parametric regimes in the
space of the coupling strength, natural frequency spread, time
delay, external forcing strength and external forcing frequency.
Using both numerical and analytical methods (wherever possible)
we obtain detailed bifurcation diagrams to mark the regions of
stability of driven synchronized solutions. We also examine the
nature of the forced amplitude response and its frequency
dependence. The system is found to display a nonlinear response
on certain critical contours in the space of the coupling strength
and time delay. Time delay plays an important role in determining
the driven frequency response and appears to offer a novel tuning knob for
controlling the system response over a wide range of
frequencies.\\

The paper is organized as follows. In Section \ref{SEC:model} we write down
the model equations for the time delay coupled driven system and
also recapitulate briefly the salient collective properties of the
non-driven, no-delay system that was studied in detail 
by Aronson, et al. \cite{AEK:90}. 
In this section we also define the particular kind of driven synchronized 
response of the coupled system that is the object of our study in the presence
of delayed coupling. We further derive the eigenvalue equation
to find the linear stability of these solutions.
In Section \ref{SEC:nodelay} we first study these driven synchronized solutions 
in the absence of time delay. 
Analytical curves describing the boundaries of the synchronized
solutions are presented. We find that when the synchronized state loses its stability,
the average frequency $<\dot\theta>$ could either grow continuously
or make a finite jump depending on the driving strength. 
The scaling behavior of the synchronized responses as a function of $F$
is studied for a special case of resonant driving ($\bar\omega=\Omega$).
Finite dispersion among the oscillators' frequencies facilitates the 
synchronization as shown by the full bifurcation diagrams.
In Section \ref{SEC:delay}
the driven synchronization mechanism is studied in the presence of time delay.
Analytic formulae of the
boundaries of the synchronized state in $(\tau,K)$ space are obtained.
The nonlinear nature of the driven response on these curves is elucidated
and its possible implications discussed.
For finite dispersion the corresponding bifurcation diagrams, and
frequency jumps across the boundaries of the stable and unstable 
synchronization regions are presented. As a function of $\tau$, the 
frequency jumps are found to scale quadratically. In Section \ref{SEC:large}
we briefly present numerical results on driven synchronization of a large number of coupled 
oscillators. Section \ref{SEC:con} summarizes our results and makes some concluding remarks
about applications and future directions of research.\\

\section{Model equations}
\label{SEC:model}
\begin{figure}
\centerline{\scalebox{1.0}{\includegraphics{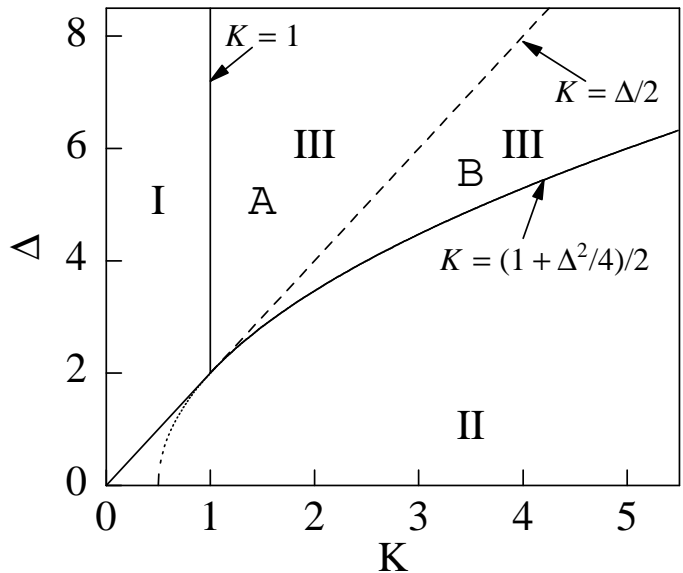}}}
\caption{\label{FIG:aronsonbif}%
Bifurcation diagram of Eqs.~(\ref{EQN:Z1Ftau}-\ref{EQN:Z2Ftau})
for $\tau=0$, and $F=0$.
}
\end{figure}

We begin by writing down the model equations for our
driven coupled system,
\be
\label{EQN:Z1Ftau}
\dot{Z}_1 = (1+i\omega_1 - \left|Z_1\right|^2)Z_1
            + K [Z_2(t-\tau) - Z_1] + F {\mathrm e}^{i \Omega t},
\ee
\be
\label{EQN:Z2Ftau}
\dot{Z}_2 = (1+i\omega_2 - \left|Z_2\right|^2)Z_2
            + K [Z_1(t-\tau) - Z_2] + F {\mathrm e}^{i \Omega t},
\ee
where $Z_j (=x_j + i y_j = r_j {\mathrm e}^{i\theta_j})$
is the complex evolution variable,
$\omega_1$, and $\omega_2$
are the intrinsic frequencies of the oscillators,
$K (>0) $ is the mutual coupling strength, $\tau (\ge 0)$ 
is the amount of time delay  in the coupling mechanism,
and both oscillators are acted upon by a uniform periodic force,
$F {\mathrm e}^{i \Omega t}$. Let us define two useful frequency parameters:
$\bar{\omega} = (\omega_1 + \omega_2)/2$, the mean frequency, and
$\Delta = \left|\omega_1-\omega_2\right|$ the natural frequency mismatch (spread)
between the two oscillators. The above coupled system without any external driving (i.e.
for $F = 0$) has been the subject of detailed studies in the past. We briefly
recapitulate the salient collective features of the undriven system. For $\tau=0$,
as demonstrated by Aronson {\it et al.} \cite{AEK:90} the system essentially shows
three distinct kinds of behavior:
(i) phase-locked (synchronized) limit cycle solutions with both oscillators oscillating
at $\omega = \bar{\omega}$, amplitudes
$r_1^2 = r_2^2 = 1-K+\sqrt{K^2-\Delta^2/4}$,
the phases $\theta_1 = \bar{\omega} t - \alpha/2$,
$\theta_2 = \bar{\omega} t + \alpha/2$ with
$\alpha = \sin^{-1}(\Delta/2K)$, (ii) asynchronous (phase
drift) behavior where each oscillator behaves nearly
independent of the other and maintains its own natural
frequency, (iii) collapse of the limit cycle amplitude to zero
(amplitude death). A composite phase diagram in the $K-\Delta$ space
delineating the parametric regimes for the occurrence of these
collective states has been provided by Aronson {\it et al.} \cite{AEK:90} 
and we reproduce their diagram as Fig.~\ref{FIG:aronsonbif} here.
The phase drift behavior (region I) occurs for all values of $\Delta>0$ as long as
the coupling strength is weak: $K < \mathrm{min} \{2K, 1\}$.
For small values of $\Delta$ phase locked behavior appears
after the coupling strength crosses a threshold value (region
II). When both the frequency disparity ($\Delta$)
and the coupling strengths are large (i.e. in the region
$1 < K < (1+\Delta^2/4)/2$, and $\Delta > 2$) amplitude death
occurs (region III). The transition from amplitude death to
phase locked behavior across the boundary of regions III and
II is in the nature of a Hopf bifurcation. In the presence of
time delay, as earlier shown by us
\cite{RSJ:98,RSJ:99}, some significant modifications
occur in the collective dynamics of the system and in the
nature of the phase diagram. For example, the collective
frequency of the phase locked state is no longer at $\bar{\omega}$
but at a frequency which has a dependence on the time delay parameter
$\tau$. In fact this dependence is in the form of a
transcendental relation giving rise to the existence of multiple frequency
locked states. Another significant change is in the topology
of the amplitude death region which can now extend down to the
$\Delta=0$ axis for some values of $\tau$ and also show
multiple-connectedness. In the presence of time delay,
identical oscillators can phase lock to either in-phase
or anti-phase locked states. Some of these delay induced
properties have also been
verified experimentally \cite{HFRPO:00,RSJexp:00}.\\

We now wish to examine the driven response
of the coupled system discussed above. Note that for finite $F$ the
trivial state ($Z_1 = Z_2 = 0$) is no more a solution of
Eqs.~(\ref{EQN:Z1Ftau}-\ref{EQN:Z2Ftau}). Thus the state of
amplitude death no longer exists. However the parametric region
corresponding to the death state can now sustain oscillations that
are externally driven and the nature of this driven response
and the domain of stability of these oscillations are the
object of our study.

Our study is focused on a particular class of driven solutions
in which each of the coupled oscillators oscillates with
the external frequency. In other words the oscillators are
not only synchronized with each other but also with the
external frequency.
The synchronization is restricted only to the frequency
and there can be a finite phase difference between
the oscillations of the two oscillators as well as with
the phase of the driving force.
These particular solutions can therefore be expressed in the
form,
\be 
\label{EQN:ftausol} Z_{1,2} (t) = R_{1,2} {\mathrm e}^{i
\theta_{1,2}}, 
\ee 
and $\theta_{1,2} = \Omega t + \alpha_{1,2}$,
where $R_{1,2}$, $\Omega$, and $\alpha_{1,2}$ are all real constants.

Substituting these solutions (\ref{EQN:ftausol})
in (\ref{EQN:Z1Ftau}) and (\ref{EQN:Z2Ftau}),
and separating the real and imaginary parts,
we arrive at four transcendental equations that determine the
amplitudes $R_{1,2}$ and the phases $\alpha_{1,2}$ for a given set of
$F$, $\Omega$, $K$, and $\tau$:
\bea
\label{EQN:ZFTR1}
(1-K-R_1^2)R_1 + K R_2 \cos A_+ + F \cos\alpha_1 = 0, \\
\label{EQN:ZFTT1}
(\omega_1-\Omega) R_1 + K R_2 \sin A_+ - F \sin\alpha_1 = 0, \\
\label{EQN:ZFTR2}
(1-K-R_2^2)R_2 + K R_1 \cos A_- + F \cos\alpha_2 = 0, \\
\label{EQN:ZFTT2}
(\omega_2-\Omega) R_2 + K R_1 \sin A_- - F \sin\alpha_2 = 0.
\eea
where $A_\pm=-\Omega\tau\pm(\alpha_2-\alpha_1)$.
Notice that the frequency of the system is specified by the
external driving frequency. The above set of equations can
possess a large number of roots (oscillatory states)
particularly when $\tau$ is finite. Except for simple limiting cases
(e.g. $\tau = 0$), these solutions can only be obtained numerically.
In addition we also need to know the region of
stability of these solutions in the $(\Delta,K)$ plane.
Following the standard procedure of linearizing around the equilibrium
solutions, it is easy to write down the
stability matrix for the above solutions as,
\be
\label{EQN:matrixM}
M = \left[
\matrix{
p_1 + i\tilde{\omega}_1   & - {R_1}^2 E_1      &  B C              & 0   \cr
-{{R_1}^2}/E_1  & p_1 - i \tilde{\omega}_1 &  0              & B/C   \cr
B C                     & 0                & p_2 + i\tilde{\omega}_2 & - {R_2}^2 E_2\cr
0                     & B/C             & -{{R_2}^2}/E_2  & p_2 - i\tilde{\omega}_2
}
\right]
\ee
where $p_{1,2} = 1 - K - 2 R_{1,2}^2$,
$\tilde{\omega}_{1,2}  = \omega_{1,2} - \Omega$,
$B = K {\mathrm e}^{-\lambda \tau}$,
$C = {\mathrm e}^{-i \Omega \tau}$,
and
$E_{1,2} = {\mathrm e}^{i 2 \alpha_{1,2}}$.
The corresponding eigenvalue equation is then given by,
\be
\label{EQN:eval2}
L^4 + a_3 L^3 + a_2 L^2 + a_1 L + a_0 = 0,
\ee
where $\lambda$ is the eigenvalue,
$L = 1 - K - \lambda$,
$a_3 = -4\,{R_1}^{2}-4\,{R_2}^{2}$,
$a_2 = -{\frac {{B}^{2}}{{C}^{2}}}
+ {\tilde{\omega}_1}^{2} + 3\,{R_1}^{4} +
{\tilde{\omega}_2}^{2} - {C}^{2}{B}^{2} +
3\,{R_2}^{4} + 16\,{R_1}^{2} {R_2}^{2}$,
$a_1 =
2\,{C}^{2}{B}^{2}{R_2}^{2} - 12\,{R_1}^{2}{R_2}^{4} +
2\,{C}^{2}{B}^{2}{R_1}^{2} -
{\frac {i {B}^{2}\tilde{\omega}_2} {{C}^{2}}} +
2\,{\frac {{B}^{2}{R_2}^{2}}{{C}^{2}}} -
4\, {\tilde{\omega}_1}^{2}{R_2}^{2} -
{\frac {i \tilde{\omega}_1\,{B}^{2}} {{C}^{2}}}-
4\,{R_1}^{2}{\tilde{\omega}_2}^{2} +
i {C}^{2}{B}^{2} \tilde{\omega}_2 +
2\,{\frac {{R_1}^{2}{B}^{2}}{{C}^{2}}} +
i {C}^{2} {B}^{2}\tilde{\omega}_1 -
12\,{R_1}^{4}{R_2}^{2}$,
and
$a_0 =
{\tilde{\omega}_1}^{2}{\tilde{\omega}_2}^{2} -
4\,{C}^{2}{B}^{2}{R_1}^{2}{R_2}^{2} + {B}^{4} -
2\,i {C}^{2}{B}^{2}\tilde{\omega}_1\,{R_2}^{2} -
2\,i {C}^{2}{B}^{2}{R_1}^{2}\tilde{\omega}_2 +
2\,{\frac {i {R_1}^{2}{B}^{2}\tilde{\omega}_2}{{C}^{2}}} +
{\frac {\tilde{\omega}_1\,{B}^{2}\tilde{\omega}_2}{{C}^{2}}}+
2\,{\frac {i \tilde{\omega}_1\,{B}^{2}{R_2}^{2}}{{C}^{2}}} -
4\,{\frac {{R_1}^{2}{B}^{2}{R_2}^{2}}{{C}^{2}}} -
{\frac {E2\,{R_1}^{2}{B}^{2}{R_2}^{2}}{E1}} +
3\,{R_1}^{4}{\tilde{\omega}_2}^{2} +
3\,{\tilde{\omega}_1}^{2}{R_2}^{4} +
{C}^{2}{B}^{2}\tilde{\omega}_1\,\tilde{\omega}_2 -
{\frac {E1\,{B}^{2}{R_1}^{2}{R_2}^{2}}{E2}} +
9\,{R_1}^{4}{R_2}^{4}$.\\

In the following sections we will solve the set of equations
(\ref{EQN:ZFTR1} - \ref{EQN:ZFTT2})
in different limits to obtain various synchronized states and use equation
(\ref{EQN:eval2}) to determine
their domains of stability along with a discussion on other possible
solutions admitted by the system.

%
%
\section{Nature of Synchronization in the absence of time delay}
\label{SEC:nodelay}
In this section we investigate the effect of force on 
the coupled system in the absence of time delay.
We start our analysis by considering the 
case of identical oscillators.
When the oscillators are identical, i.e. $(\omega_1 = \omega_2 = \omega)$,
and undriven, i.e. $F=0$, the system admits two kinds of solutions: 
(i) in-phase solutions
characterized by $Z_{1} = Z_{2} = Z$,
and (ii) anti-phase solutions characterized by
$Z_{1}=-Z_{2} = Z$.
We will study here the in-phase solutions under finite force.
Under the assumption of the in-phase solutions, the study of
Eqs.~(\ref{EQN:Z1Ftau}) and (\ref{EQN:Z2Ftau})
becomes identical to studying a single oscillator with
a delayed linear feedback and constant external forcing.
Thus, it is sufficient to
focus on the following simple equation:
\be
\label{EQN:ZFident}
\dot{Z}(t) = (1 +i\omega - \left|Z(t)\right|^2)Z(t) + F {\mathrm e}^{i\Omega t}
\ee
or under the transformation $Y(t) = Z(t) {\mathrm e}^{-i \Omega t}$,
\be
\label{EQN:YFident}
\dot{Y}(t) = (1 +i\tilde{\omega} - \left|Y(t)\right|^2)Y(t) + F,
\ee
where $\tilde{\omega} = \omega-\Omega$.
Such a set of equations in the context a single relaxation 
oscillator under an external force was earlier studied
by Guckenheimer and Holmes \cite{GH:83book},
and as shown by them, the system admits two kinds of
solutions: one a stationary state of the above equation that corresponds
to a periodic solution of Eq.~(\ref{EQN:ZFident}), and the other, a periodic
solution that corresponds to a solution of Eq.~(\ref{EQN:ZFident}) with two
frequencies. Both these solutions in our present context would correspond to
the in-phase oscillations of the two identical oscillators represented by
Eqs.~(\ref{EQN:Z1Ftau}-\ref{EQN:Z2Ftau}). The stable fixed point of
Eq.~(\ref{EQN:YFident}) would correspond to a synchronized state of the
oscillators that is also synchronized  with the external frequency, whereas
the periodic solution of  Eq.~(\ref{EQN:YFident}) would represent
a synchronized state of the
oscillators that is not synchronized with the external frequency.

The stationary solution of Eq.~(\ref{EQN:YFident}) is the 
synchronized solution of the system.
Substituting $Y = R {\mathrm e}^{i\alpha}$ in
Eq.~(\ref{EQN:YFident}) and separating the real and imaginary parts
one arrives at the following two equations:
$ (1-R^2) R + F \cos \alpha   = 0,$  and
$\tilde{\omega} R - F \sin \alpha = 0, $
which can be algebraically
simplified to arrive at the following expressions
for determining $R$ and $\alpha$:
\bea
\label{EQN:Req2}
R^6 - 2 R^4 + (1+{\tilde{\omega}}^2) R^2 - F^2 = 0, \\
\label{EQN:alphaeq2}
\alpha = \cases{
\sin^{-1} \frac{\tilde{\omega} R}{F} & ~~if~~ $1 < R^2$,\cr
\pi-\sin^{-1}\frac{\tilde{\omega} R}{F} & ~~if~~ $1 \ge R^2$.
}
\eea
Eq.~(\ref{EQN:Req2}) produces a curve 
$R = \gamma(\tilde{\omega},K,\Omega\tau,F)$
which can be multiple-valued. The number of real roots
for ${R}^2$ now ranges from one to three.
This number is decided by the following factor:
\bea
G = \frac{1}{27}\left[B^3 - B^2-9 F^2 B +
\frac{27}{4} F^4 + 8 F^2\right], \nonumber
\eea
where $B=1+\tilde{\omega}^2$.
There will be one real solution if $G > 0$, three real solutions
either if $G=0$ (in which case at least two of them are identical)
or if $G < 0$.
In particular, for $\tilde{\omega} = 0$ and $\tau = 0$, there will be
one real solution when $F>\frac{2}{3\sqrt{3}}$,
two solutions at the points $F = 0$ and $F = \frac{2}{3\sqrt{3}}$,
and three solutions when $0 < F < \frac{2}{3\sqrt{3}}$.

The stability of these fixed points must be determined starting from
matrix ~(\ref{EQN:matrixM}) by inserting $\omega_1 = \omega_2 = \omega$.
This in the present case simplifies to the following
four equations:
\bea
\label{EQN:wwFFeval1}
\lambda = 1- 2 {R}^2 \pm \left\{{R}^4-\tilde{\omega}^2\right\}^{1/2}, \\
\label{EQN:wwFFeval2}
\lambda = 1- 2 K - 2 {R}^2 \pm \left\{{R}^4-\tilde{\omega}^2\right\}^{1/2}.
\eea
We next examine these equations to determine the regions of stability
of the fixed points.
The curves that define
the stability region of the synchronized state are similar in form
to those discussed by Guckenheimer and Holmes \cite{GH:83book} 
in their study of
a single driven Hopf bifurcation oscillator.
However, we would now have an additional set of 
three curves arising due to the higher dimensionality of
the problem. (All the critical curves $\Gamma_1$, $\ldots$, $\Gamma_6$
are derived in the Appendix.)
\begin{figure}
\centerline{\scalebox{0.8}{\includegraphics{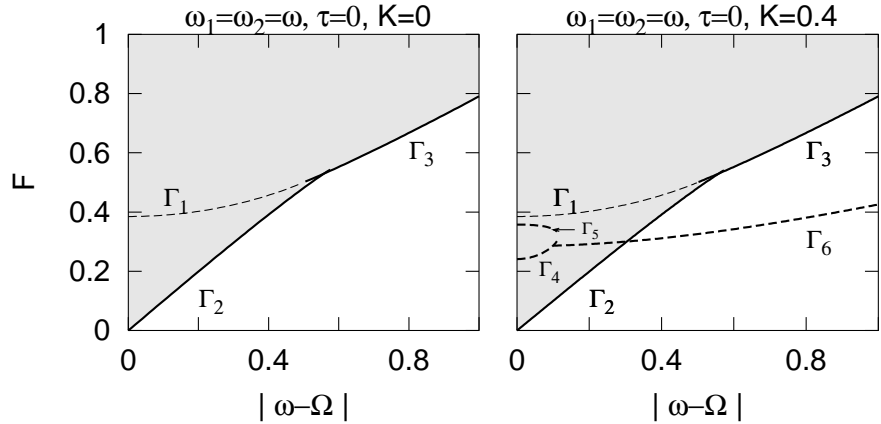}}}
\caption{\label{FIG:fw1}%
Bifurcation diagram of the in-phase solutions of identical oscillators with no delay.
The shaded region represents the stability region of $Z_{1,2} = R e^{i(\Omega t+\alpha)}$.
}
\end{figure}
In Fig.~\ref{FIG:fw1} all these critical curves
are plotted for two different values of $K$.
Notice that for $\tilde{\omega}=0$, the eigenvalues
become $\{1-R^2, 1-3 R^2\}$.
So the branch of the fixed points ($\gamma$)
falling above $R_{1,2} = 1$ is stable for all $F$, and that below is unstable.
This lower boundary of stability is given by $\Gamma_2$.
The two unstable branches arising due to multiplicity of $\gamma$
merge on and disappear above $\Gamma_1$. At large values of $\tilde{\omega}$,
the lower boundary of stability region is $\Gamma_3$.
When $1/\sqrt{3} < \left|\tilde{\omega}\right| < 1/2$, the curve
$\gamma$ acquires stability in its multi-valued region leading to
bistability of fixed points.
The shaded region in Fig.~\ref{FIG:fw1} thus represents the stability region
in which the system has at least one stable fixed point.

The stability analysis also produced an additional set of three curves $\Gamma_4$,
$\Gamma_5$ and $\Gamma_6$ which, for $K = 0$ merge with the curves
$\Gamma_1$, $\Gamma_2$ and $\Gamma_3$ respectively, and do not exist for $K>0.5$.
These indicate the intersections or mergers of other possible solutions
of the system with the in-phase solutions studied above.
A closed form for these solutions is not possible. But
all such existing solutions can, in principle,
be determined from the algebraic
relations $\dot{Z}_{1.2} =0$ by setting $\omega_1 = \omega_2 = \omega$.

%
%
\subsection{Amplitude response and frequency jumps}
We now study the response of the system as a function of $F$.
The response of the system in the region of stable synchronized solution
is given by Eq.~(\ref{EQN:Req2}). In particular notice that since the
basic oscillator has the limit cycle solution with a fixed amplitude of unity,
the response is always linear for small amplitudes as a function of $F$.
And at large amplitudes, the response is nonlinear: $R\propto F^{1/3}$.
The in-phase oscillations will be locked to the external frequency with
a phase difference $\alpha = 0$ when $\tilde\omega=0$ 
(i.e. when they are driven resonantly),
and with a phase difference $\alpha = \pi$ on the curve $F = \tilde{\omega}$.

Since the stable region for small $F$ is accessible
only for $\left|\omega-\Omega\right|=0$, it is only when the system is
driven with $\omega$ that a synchronization with the external frequency occurs
for small $F$. Otherwise the system always goes to a two frequency state.

\begin{figure}
\centerline{\scalebox{0.8}{\includegraphics{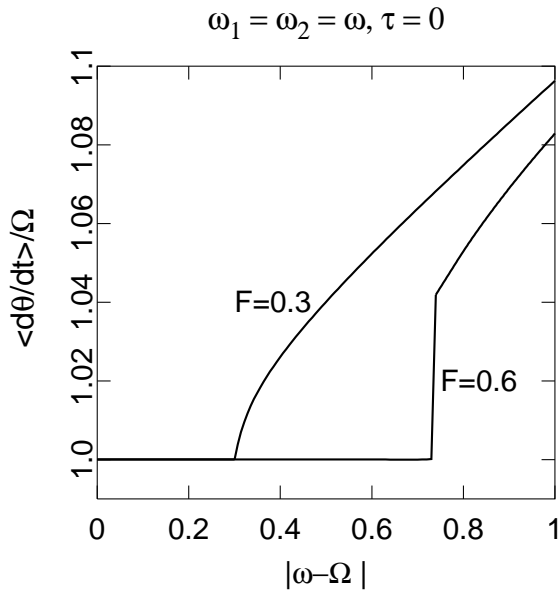}}}
\caption{\label{FIG:respavew1w2w}%
Frequency transitions for different value of $F$ as 
$\left|\omega-\Omega\right|$ is increased.
}
\end{figure}

It is now left to determine the nature of the system in the region where
the symmetric in-phase solution is unstable. Of particular interest is
the average frequency of the system when compared to the frequency of
the driving. In Fig.~\ref{FIG:respavew1w2w} the quantity
$<\dot{\theta}>/\Omega$
is plotted as a function of $\left|\omega-\Omega\right|$ while keeping $\Omega = 10$.
The plot shows that this quantity has a finite jump across $\Gamma_3$ and no jump
across $\Gamma_2$ which is consistent with the fact that on the critical curve $\Gamma_3$
there is a finite imaginary quantity of the eigenvalue, where as on $\Gamma_2$, the
imaginary quantity is zero (and hence the frequency is equal to the external frequency).
In fact the jump $(J)$ is directly related to the imaginary quantity on 
the critical curves
across which the transition takes place:
\bea
 J = \sqrt{(\omega-\Omega)^2-\frac{1}{4}} \nonumber
\eea

Finally it should be remarked that for identical oscillators,
region I (see Fig.~\ref{FIG:aronsonbif})
corresponding to incoherent states (for $F=0$) does not exist.
Hence the above transition curve is the only relevant one.

%
%
\subsection{Effect of finite dispersion}
Finite dispersion facilitates the synchronization mechanism
as we describe in this section using bifurcation diagrams.
As seen in Fig.~\ref{FIG:aronsonbif} regions $A$ and $B$ 
together make up the amplitude death region in the absence
of an external force. The region is bounded by the curves
$K = 1$ and $K = \gamma(\Delta) = \frac{1}{2}(1+\Delta^2/4)$.
The eigenvalues that determine the stability of this steady state
are given by $\lambda = 1 - K \pm i \bar{\omega} \pm \sqrt{K^2-\Delta^2/4}$.
The regions $A$ and $B$ are separated by the curve $K = \Delta/2$.
Note that the system in the death state shows only one collective
frequency $f = \bar{\omega}$, which occurs in region $B$.
In region $A$, i.e, when $K<\Delta/2$, the oscillators
spiral to the death state with independent frequencies:
$f_{1,2} = \bar{\omega}\pm\sqrt{\Delta^2/4-K^2}$.
In the presence of $F$, the external frequency
interacts with $f_{1,2}$ and $f$ in the regions $A$ and $B$ respectively.
Since the system offers damping to these modes, the system as a whole
acts like a damped nonlinear oscillator with these characteristic frequencies.
Thus the frequency of oscillation under forcing in both the regions
is that of the external driver. The region of stability of this driven synchronization
(i.e. the solution $Z_{1,2} = R_{1,2} {\mathrm e}^{i(\Omega t + \alpha_{1,2})}$)
will have to be self consistently determined using the 
eigenvalue equation described in the previous section.
Hence in the presence of $F$, the region of death now transforms into
an oscillatory region in which both the oscillators synchronize with
one another and also synchronize with the external frequency.
The actual region of stability of the synchronized solution
envelops the earlier death region. Before we present
the stability region of this solution, we make some observations
about the response of the individual oscillators. Without loss
of generality, let us assume $\omega_1 > \omega_2$.
If $\Omega > \bar{\omega}$, then $R_1 > R_2$.
If $\Omega = \bar{\omega}$, then $R_1 = R_2$.
And if $\Omega < \bar{\omega}$, then $R_1 < R_2$.
At the resonant driving, the sum of the phases of the oscillators
with respect to the external driver will be $0$.
i.e. $\alpha_1+\alpha_2 = 0$.

\begin{figure}
\centerline{ \scalebox{0.4}{\rotatebox{0}{\includegraphics{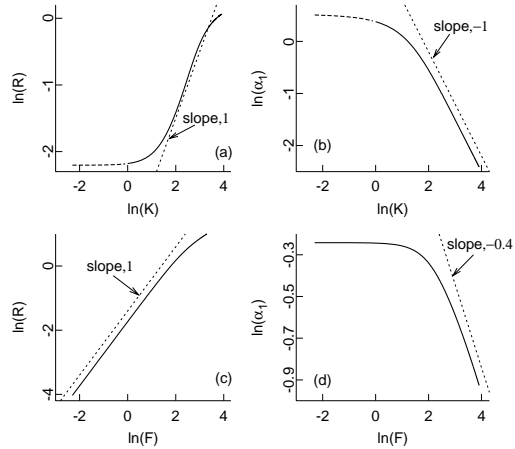} } } }
\caption{\label{FIG:Ralpha_KFlog}%
Dependence of $R$ and $\alpha_1$ on $K$ and $F$. (a) and (b) are plotted
for $F=1$, and (c) and (d) are plotted for $K=5$. The other parameters
for all the plots are $\bar{\omega}=10$, $\Delta=18$, $\Omega=10$, and
$\tau = 0$.
}
\end{figure}

We now briefly describe the solutions when the oscillators are
driven resonantly, i.e. when $\Omega = \bar{\omega}$.
At resonant driving, it is possible to arrive at a simpler set of
algebraic equations to determine the amplitudes and the phase of
this oscillatory state. Substituting $R_1 = R_2 = R$
and $\alpha_2 = -\alpha_1$
in Eqs.~(\ref{EQN:ZFTR1}-\ref{EQN:ZFTT2}),
the following two relations can be written down
that describe the amplitude and the phase of the system.
$$
R = {F \sin\alpha_1} / [\tilde{\omega}_1 - K \sin(2\alpha_1)],
$$
where $\alpha_1$ is determined from the following functional relation:
$$
g( \alpha_1 ) =  F^2 \sin^3\alpha_1
   - (\tilde{\omega}_1 - K \sin(2\alpha_1))^2 
 [ 1 - K + K \cos(2\alpha_1)] \sin\alpha_1
    - (\tilde{\omega}_1-K \sin(2\alpha_1))^3 \cos\alpha_1 = 0.
$$
The response grows linearly with force. The dependence of $\alpha$
however is not apparent from the equation.
Notice that the resonant point
does not necessarily fall on the critical point of Hopf-bifurcation.
The response of the oscillators and the phases are shown 
in Fig.~\ref{FIG:Ralpha_KFlog} as a function of both
$K$ and $F$.
For small $F$, $R$ grows linearly, and for large $F$, it grows sub-linearly
according to $R\propto F^{1/3}$.
The scaling of the amplitude as a function of $F$ in a
linear fashion is in accordance with the fact that inside the region of
death state, the driven oscillator shows a linear response.

\begin{figure}
\centerline{ \scalebox{0.5}{\rotatebox{0}{\includegraphics{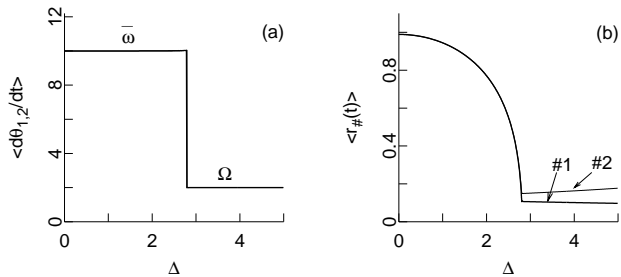} } } }
\caption{\label{FIG:ave_vs_delta_tau0}%
The time averaged quantities $<\dot{\theta}_{1,2}>$ and $<r_{1,2}>$
plotted in (a) and (b) respectively.
}
\end{figure}

The coupled system shows an interesting response as a function of $K$.
Increasing $K$ will sweep the stable synchronized region. For small
$K$, the system shows incoherence, which is marked with dashed lines.
The synchronized response loses stability at large values $K$
in a Hopf bifurcation. Close to this critical boundary, the phase of
of the first oscillator as shown in the figure damps as $1/F$.
However, the amplitude has a very nonlinear behavior as a function of $F$.
It can be considered as linear only in the middle of the stability region.
Note, however, that since the region encompasses a
Hopf bifurcation curve, $\gamma(\Delta)$,
the response of each of the oscillators
becomes nonlinear and proportional to $F^{1/3}$ on this curve.
This characteristic nonlinear response at the Hopf bifurcation
threshold has been invoked earlier to model the behavior
of inner hair cells in human cochlea \cite{CDJP:00,EOCHM:00}.
In the present context of two coupled oscillators,
we have numerically verified this feature.
The phase $\alpha_1$, which is a measure of the
phase difference
the first oscillator maintains with the driver, is nonzero,
and is nearly constant for all weak forces, but damps with
an approximate scaling of $F^{-0.4}$ for large values of $F$. 
The dotted curve drawn is only an indication of the power law.
This behavior was not noted in literature
before and may have some practical applications.
\begin{figure}
\centerline{ \scalebox{0.8}{\rotatebox{0}{\includegraphics{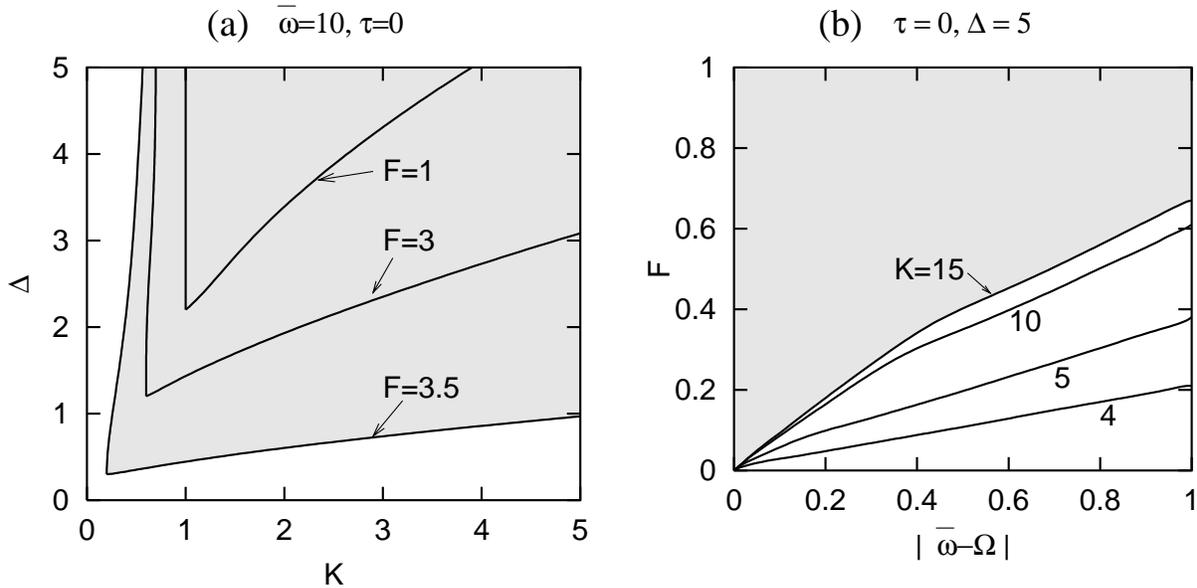} } } }
\caption{\label{FIG:CritKdeltanodelay}%
Stability regions (shaded) of the synchronized solutions
$Z_{1,2} = R_{1,2} {\mathrm e}^{i(\Omega t + \alpha_{1,2})}$ when
$\tau = 0$
(a) for various $F$ when $\bar{\omega} = 10$,
(b) in $(F,\left|\tilde{\omega}\right|)$.
}
\end{figure}
The coefficients of the eigenvalue equation (\ref{EQN:eval2}) reveal that
the stability is a function of the product term $\Omega\tau$. 
Thus the stability region in the absence of time delay is independent of $\Omega$ and only
depends on $F$. We now present the stability regions of this synchronized
state in the parametric space of $(K,\Delta)$ for different values of $F$
in Fig.~\ref{FIG:CritKdeltanodelay}(a). 
For weak forcing, the oscillators can be moved out of 
the synchronized regime by strongly coupling them. But
as the forcing strength increases, much stronger
or much weaker coupling is required to destabilize
the synchronized state. At weaker couplings, it is the
dispersion ($\Delta$) that is responsible for instability,
and at stronger couplings, the system is more likely to get
locked to its average frequency than to the external frequency.
The relation between $F$ and $\left|\bar\omega-\Omega\right|$ is
shown in Fig.~\ref{FIG:CritKdeltanodelay}(b).
At larger values of $K$, a strong force is required to synchronize the
system to the external frequency.
When these solutions become unstable, there are two other possible
solutions that are similar in character to the synchronized, and
the incoherent states of the non-driven system. For any of the parameters
corresponding to the bifurcation diagrams in
Fig.~\ref{FIG:CritKdeltanodelay}(a), we note that the system's average
frequency is locked
to the average frequency $\bar{\omega}$ of the oscillators
for small value of $\Delta$.
A numerical plot of the averaged frequency is shown in
Fig.~\ref{FIG:ave_vs_delta_tau0}(a) which reveals that the transition from
one coherent state to another is discontinuous. The nature of bifurcation
at this transition is supercritical Hopf as can be seen from 
Fig.~\ref{FIG:ave_vs_delta_tau0}(b).

%
%

\section{Nature of Synchronization in the presence of time delay}
\label{SEC:delay}
In this section we investigate effect of time delay on the forced synchronization of
the coupled system.
We again start our analysis by considering the instructive 
case of identical oscillators.
We will study here the in-phase solutions under finite force.
As in the previous section, 
under the assumption of the in-phase solutions, the study of
Eqs.~(\ref{EQN:Z1Ftau}) and (\ref{EQN:Z2Ftau})
becomes identical to studying a single oscillator with
a delayed linear feedback and constant external forcing.
Thus, it is sufficient to
focus on the following equation:
\be
\label{EQN:ZFtauident}
\dot{Z}(t) = (1 - K +i\omega - \left|Z(t)\right|^2)Z(t) + K Z(t-\tau) + F {\mathrm e}^{i\Omega t}
\ee
or under the transformation $Y(t) = Z(t) {\mathrm e}^{-i \Omega t}$,
\be
\label{EQN:YFtauident}
\dot{Y}(t) = (1 - K +i\tilde{\omega} - \left|Y(t)\right|^2)Y(t)
              + K {\mathrm e}^{- i \Omega \tau} Y(t-\tau) + F,
\ee
where $\tilde{\omega} = \omega-\Omega$.
Substituting $Y = R {\mathrm e}^{i\alpha}$ in
Eq.~(\ref{EQN:YFtauident}) and separating the real and imaginary parts
one arrives at the following two equations:
\bea
(1-K-R^2) R + KR \cos( \Omega \tau ) + F \cos \alpha   = 0, \nonumber \\
\tilde{\omega} R - K R \sin( \Omega \tau )- F \sin \alpha = 0, \nonumber
\eea
which can be algebraically
simplified to arrive at the following expressions
for determining $R$ and $\alpha$:
\bea
\label{EQN:Req2tau}
R^6 + a_2 R^4 + a_1 R^2 - F^2 = 0, \\
\label{EQN:alphaeq2tau}
\alpha = \cases{
\sin^{-1} \frac{b R}{F} & ~~if~~ $c < R^2$,\cr
\pi-\sin^{-1}\frac{b R}{F} & ~~if~~ $c \ge R^2$.
}
\eea
where $b = \tilde{\omega}-K\sin(\Omega\tau)$,
$a_1 = b^2 + c^2$.
$a_2 = -2 c$, and $c = 1-K+K\cos(\Omega\tau)$.
Eq.~(\ref{EQN:Req2tau}) produces a curve $R = \gamma(\tilde{\omega},K,\Omega\tau,F)$
which can be multiple-valued. The number of real roots
for ${R}^2$ now ranges from one to three.
The stability of these solutions that can be
determined from $M$ are governed by the following eigenvalue  
equations,
\be
\label{EQN:wwFFevaldelay}
\lambda  =  1 - K - 2 {R}^2 + s_k K {\mathrm e}^{-\lambda \tau} +
                 s \sqrt{{R}^4-\tilde{\omega}^2}, \\
\ee
where $s_k = \pm 1$ and $s = \pm 1$. These two signs must be taken in
all the permutations. Thus we have four transcendental equations.
It is possible to treat all these equations together.
Just as we did in the no-delay case, we consider here also two cases
(i) ${R}^2 > \left| \tilde{\omega} \right|$ and
(ii) ${R}^2 \le \left| \tilde{\omega} \right|$.
In the first case, in order to arrive at the critical curves across
which transitions of eigenvalues take place, we set $\lambda = i \beta$.
This gives the following two equations:
\bea
\label{EQN:brr1}
  1 - K - 2 {R^*}^2 + s  \sqrt{{R^*}^4- \tilde{\omega}^2} = - s_k K \cos(\beta\tau), \\
\label{EQN:brb1}
  \beta = - s_k K \sin( \beta \tau ).
\eea
Note that $\beta = 0$ is a solution of the Eq.~(\ref{EQN:brb1}) for any
value of $K$ and $\tau$. This branch corresponds to the critical curves
that existed for the case of $\tau=0$ when both $s$ and $s_k$ are
considered independently. For a second branch of solution for
Eq.~(\ref{EQN:brb1}) to exist, the necessary condition is
$K\tau > 1$.  Again from Eq.~(\ref{EQN:brb1}), it can further be
concluded that the critical values of $K\tau$ beyond which the
second branch exists are determined from the zeros of the function
$g = \cos(\sqrt{K^2\tau^2-1})+1/(s_K K \tau)$. A numerical analysis
of the equation $g = 0$ gives the critical values:
$K\tau = 4.6034$, if $s_K = 1$, and $K\tau = 1$, if $s_k = -1$.
A set of critical curves for the second and higher branches
are obtained by explicitly considering $s = \pm 1$ of the newly
generated branches. Also note that the response now is a function of
$\tau$. Hence the stability can be affected by the $\beta = 0$ branch
itself.
An equation for $R^2$ can be written down by eliminating $\beta$
as follows:
\bea
\beta = \left\{ K^2 - \left[ 1 - K - {R^*}^2
         + s \left\{{R^*}^4- \tilde{\omega}^2\right\}^{1/2}\right]^2 \right\}^{1/2}, \\
{R^*}^2 = \frac{1}{2} \left(1-K + s_k K \cos( \beta \tau )
                   + s \sqrt{{R^*}^4- \tilde{\omega}^2}\right).
\eea
The critical curves in $(F, \tilde{\omega})$ plane cannot in general
be written down
in a closed form as we did for the case of $\tau= 0$.
The conditions that exist on $K$, and $\tilde{\omega}$
must now be determined using the two transcendental equations above.
These equations can be used along with Eq.~(\ref{EQN:Req2tau}) to
numerically plot the critical curves in $(F,\tilde{\omega})$ plane by
setting $R^2 = {R^*}^2$. We can easily identify that the curves that are
counterparts of $\Gamma_1$ and $\Gamma_2$ of
the no-delay case are generated here
by setting $s_k = 1$.
The critical curves corresponding to $\beta = 0$ can, however,
be determined in closed form.
Following a similar analysis as outlined 
in the Appendix, these curves are
obtained as
\bea
\Gamma_1^\prime = \left\{ F = f(\rho_+) \big|
\rho_+ = \frac{2}{3} + \frac{1}{3} \sqrt{1-3\tilde{\omega}^2} \right\},
         \omega^2 \in [0,\frac{1}{3}], \\
\Gamma_2^\prime = \left\{ F = f(\rho_-) \big|
\rho_+ = \frac{2}{3} - \frac{1}{3} \sqrt{1-3\tilde{\omega}^2} \right\},
         \omega^2 \in [0,\frac{1}{3}],
\eea
where $f(\rho) = \{\rho^3 + a_2 \rho^2 + a_1 \rho\}^{1/2}$.
These are the only critical curves that exist under this case as long as
$K\tau <1$. If this condition is violated pairs of sets of such critical
curves exist.

\begin{figure}
\centerline{\scalebox{0.8}{\includegraphics{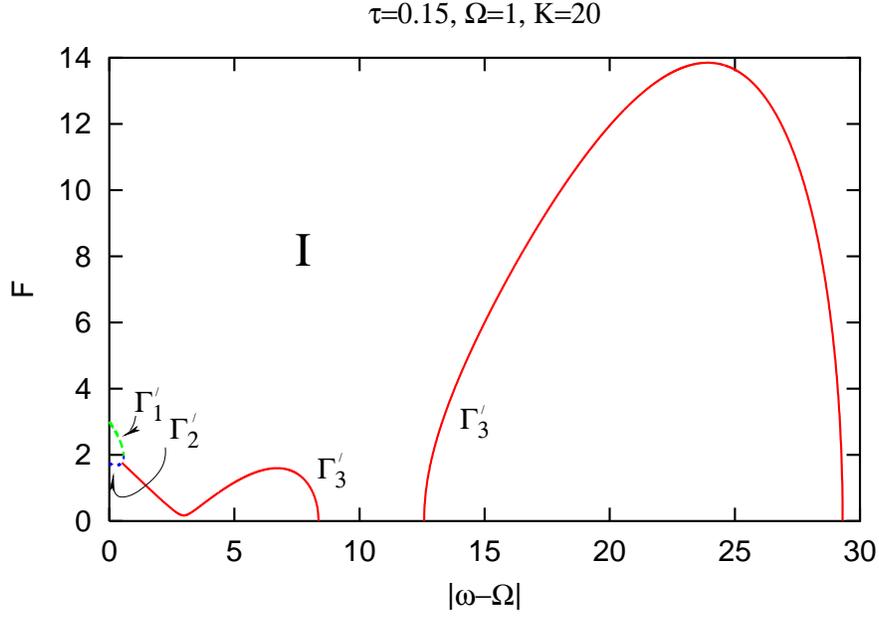}}}
\caption{\label{FIG:tau0p5fig}%
The critical curves
$\Gamma_1^\prime$,
$\Gamma_2^\prime$, and
$\Gamma_3^\prime$,
in the presence of $\tau$ at $K=20$, $\Omega=1$, and $\tau = 0.15$.
Region I represents the synchronized region.
}
\end{figure}

In the second case, i.e. when ${R}^2 \le \left|\tilde{\omega}\right|$,
on the critical curves, we again set $\lambda = i \beta$ to arrive at
the following set of equations:
\bea
\label{EQN:brr2}
  1 - K - 2 {R^*}^2 = - s_k K \cos(\beta\tau), \\
\label{EQN:brb2}
  \beta + s_k K \sin( \beta \tau ) = s \sqrt{{R^*}^4 - \tilde{\omega}^2}.
\eea
Note that since the eigenvalues occur in complex conjugate pairs,
the second of the above two equations yields the same set of curves
for both $s=\pm 1$. So there will be two classes of curves corresponding
to $s_k = \pm 1$. The above two equations can further be simplified to
yield
\bea
\beta = \beta_\pm=\sqrt{\tilde{\omega}^2-{R^*}^4} \pm \sqrt{K^2-(1-K-2 {R^*}^2)}, \\
{R^*}^2 = \frac{1}{2} \left(1-K + s_k K \cos( \beta \tau ) \right).
\eea
The second set of critical curves, $\Gamma_3^\prime$,
in $(\tilde{\omega},F)$ plane are obtained
using the above two equations along with Eq.~(\ref{EQN:Req2})
by setting $R^2 = {R^*}^2$. We can again identify that the counterpart of
the curve $\Gamma_3$ of the no-delay case is obtained here
by setting $s_k = 1$.
As $\tau$ is increased, the parameter region falling below the
curve $\Gamma_3$ becomes more and more susceptible for 
synchronization, and in fact for a range of value of $\tau$
the curve makes intersections with $F=0$ line as shown in 
Fig.~\ref{FIG:tau0p5fig}.
The region marked I is the
region of stability of the synchronized solution. At the region of
intersection of the three curves, there is also a possibility of more
than one synchronized solution coexisting.
At higher values of time delay, a stronger force is required
to make the oscillators synchronize with the external driver.
The other set of curves that are generated using the second
sign $s_k = -1$ represent intersections of the response of
the system with other unstable solutions the system possesses.
We do not discuss them here. This figure depicts one of the
important results of our present paper. In the presence of appropriate time delay
the region of forced synchronization can be extended to continuously connect
with $F=0$ axis. This raises the interesting possibility of achieving
synchronization far from resonant driving with a minimal (nearly infinitesimal) force.

The role of time delay can be more clearly appreciated by determining
the required critical curves in the plane of $(\tau,K)$.
By following the usual analysis of such characteristic equations and
properly choosing the correct signs for the cosine function above,
the critical curves in the plane of $(K,\tau)$ can easily be
derived. When ${R^*}^2 > \left|\tilde{\omega}\right|$,
the equations for $R$ and $\beta$ at criticality,
Eqs.~(\ref{EQN:brr1}) and (\ref{EQN:brb1}) can be used to derive
an equation for $\tau$ in terms of $K$:
\be
\tau_\pm = \frac{n\pi-\cos^{-1}\left(\frac{-A_\pm}{K}\right)}
              {\sqrt{K^2-A_\pm^2}},
\ee
where $A_\pm = 1 - K - 2 {R^*}^2 \pm \sqrt{{R^*}^4-\tilde{\omega}^2}$.
On both of the above two curves, $\tau_\pm$, the nature of the transition
of the eigenvalues is given by:
\be
\frac{d}{d\tau}Re(\lambda)\Big|_{Re(\lambda) = 0} > 0.
\ee
Thus these two curves in the $(\tau,K)$ space represent critical curves
across which a pair of eigenvalues makes transition to acquire positive
real parts.

In the second case, i.e. when ${R^*}^2 \le \left|\tilde{\omega}\right|$,
the valid equations are Eqs.~ (\ref{EQN:brr2}) and (\ref{EQN:brb2}).
By again following the standard methods to derive the critical curves,
we arrive at the following
set of two marginal stability curves:
\bea
\tau_1 = \frac{n\pi-\cos^{-1}\left(\frac{1-K-2{R^*}^2}{-K}\right)}
              {\sqrt{\tilde{\omega}^2-{R^*}^4} + \sqrt{K^2-(1-K-2 {R^*}^2)^2}}, \\
\tau_2 = \frac{n\pi+\cos^{-1}\left(\frac{1-K-2{R^*}^2}{-K}\right)}
              {\sqrt{\tilde{\omega}^2-{R^*}^4} - \sqrt{K^2-(1-K-2 {R^*}^2)^2}}.
\eea
The nature of transition of the eigenvalues across these critical
curves is given by:
\be
\frac{d}{d\tau}Re(\lambda)\Big|_{Re(\lambda) = 0} \cases{
                 > 0 ~~{\mathrm on }~~ \tau_1,\cr
                 < 0 ~~{\mathrm on }~~ \tau_2.}
\ee
The curves $\tau_\pm$ and $\tau_{1,2}$ completely describe the
stability transition curves in $(\tau,K)$ space for a given
set of $F$ and $\omega$. The ordering of these curves depending
on $F$ and $\omega$ will eventually
determine the curves that enclose the stable regions.

%
%
\subsection{Amplitude response and frequency jumps}
\begin{figure}
\centerline{\rotatebox{270}{\scalebox{1.0}{\includegraphics{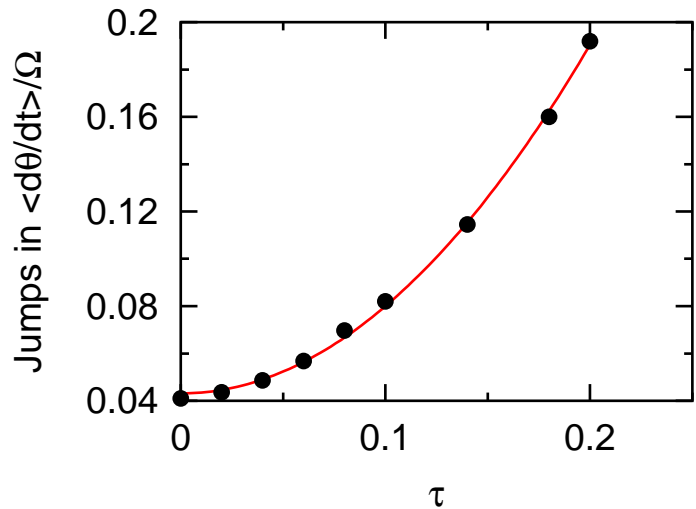}}}}
\caption{\label{FIG:jumpsvstau}%
Jumps of the average frequency across the
critical boundary separating synchronous and non-synchronous regions.
$K=0.4$, $F=0.6$, $\Omega=10$. Bullets are numerically found
values. The curve is a best fit with $3.685\tau^2+0.043$.
}
\end{figure}

We now examine the response of the system. Notice from
Eq.~(\ref{EQN:Req2tau}) that the oscillators respond linearly with $F$
for small amplitudes (i.e. when the $F^2$ term is balanced by the
$a_1 R^2$ term). For larger amplitudes the response becomes nonlinear
as the balance is provided by other terms.
However the response can
be highly nonlinear even for small amplitudes on certain
contours of the parametric space where $a_2 = 0$, and $a_1 = 0$.
Under these conditions the following two equations are obtained
(provided that $\Omega\tau \ne 2 n \pi$):
$$
\cos(\Omega\tau) = (1-K)/K,
$$
and
$$
\sin(\Omega\tau) = (\omega-\Omega)/K. 
$$
Using the above two equations we can derive the conditions
relating $K$ and $\Omega\tau$ for a given $\tilde{\omega}$ such that
the response of the system for small amplitudes has a nonlinear
character ($R\propto F^{1/3}$):
\bea
\label{EQN:conditionK}
K & = & (1+\tilde{\omega}^2)/2, \\
\label{EQN:conditionOT}
\Omega \tau & = & 2 n \pi + \cos^{-1}\left(\frac{1-\tilde{\omega}^2}{1+\tilde{\omega}^2}\right).
\eea
In the $(F,\tilde{\omega})$ plane, the corresponding contours can be
derived from Eq.~(\ref{EQN:Req2tau}) by making use of the above two relations.
We will discuss this phenomenon in more detail later.
At the boundary of stability region, the oscillators make a transition
in their average frequency from their intrinsic average frequency to that
of the driver. This discontinuity is a function of $\tau$ and is numerically
found to be quadratic in its dependence on $\tau$ as shown 
in Fig.~\ref{FIG:jumpsvstau}.

%
%
\subsection{Effect of finite dispersion}
We now study the effect of finite time delay and the external force
on non-identical oscillators. Our primary objective is to examine
the stability properties of the synchronized state
$Z_{1,2} = R_{1,2} {\mathrm e}^{i (\Omega + \alpha_{1,2})}$
which is also
synchronized with the external frequency.
We use Eqs.~(\ref{EQN:ZFTR1}-\ref{EQN:ZFTT2})
to determine the responses $R_{1,2}$, and $\alpha_{1,2}$
and then find the eigenvalue spectrum to determine their
stability. The response due to time delay shows considerable
deviation from the no-delay case. 
Under certain conditions, both the oscillators could have identical
amplitudes. Similar to our studies in the previous section, we
assume that $R_1 = R_2 = R$ and $\alpha_1 + \alpha_2 = 0$. This
ansatz leads to the condition that $\Omega\tau = \pi$. Substituting
these relations in Eqs.~(\ref{EQN:ZFTR1}-\ref{EQN:ZFTT2}),
we arrive at the following
two simplified equations for the responses:
\bea
\cos\alpha_1 & = & -\frac{R}{F} \left(1 - K - R^2 - K \cos(2\alpha_1) \right), \nonumber \\
\sin\alpha_1 & =  & \frac{R}{F} \left( \tilde{\omega}_1 + K \sin(2\alpha) \right), \nonumber
\eea
which can further be simplified to arrive at an equation for $R$
as a function of $\alpha_1$ as
\bea
R = {F \sin\alpha_1} / [\tilde{\omega}_1 + K \sin(2\alpha_1)], 
\eea
and a functional relation to determine $\alpha_1$ as
$$
h( \alpha_1 ) = F^2 \sin^3\alpha_1
   - (\tilde{\omega}_1 + K \sin(2\alpha_1))^2 
   [ 1 - K - K \cos(2\alpha_1)] \sin\alpha_1 
    - (\tilde{\omega}_1+K \sin(2\alpha_1))^3 \cos\alpha_1 = 0. 
$$
The above expression for $\cos\alpha_1$ again indicates that
for this special case the response of the oscillators is
nonlinear according to $R\propto F^{1/3}$ at large values of $F$.

\begin{figure}
\centerline{\scalebox{0.4}{\rotatebox{270}{\includegraphics{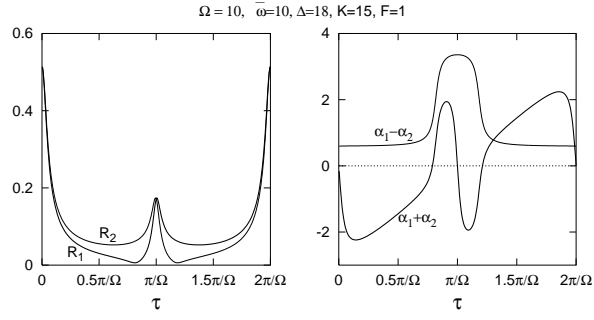} } } }
\caption{\label{FIG:respvstau}%
Amplitudes and phases of the oscillators as a function of $\tau$.
}
\end{figure}

\begin{figure} 
\centerline{\scalebox{0.8}{\includegraphics{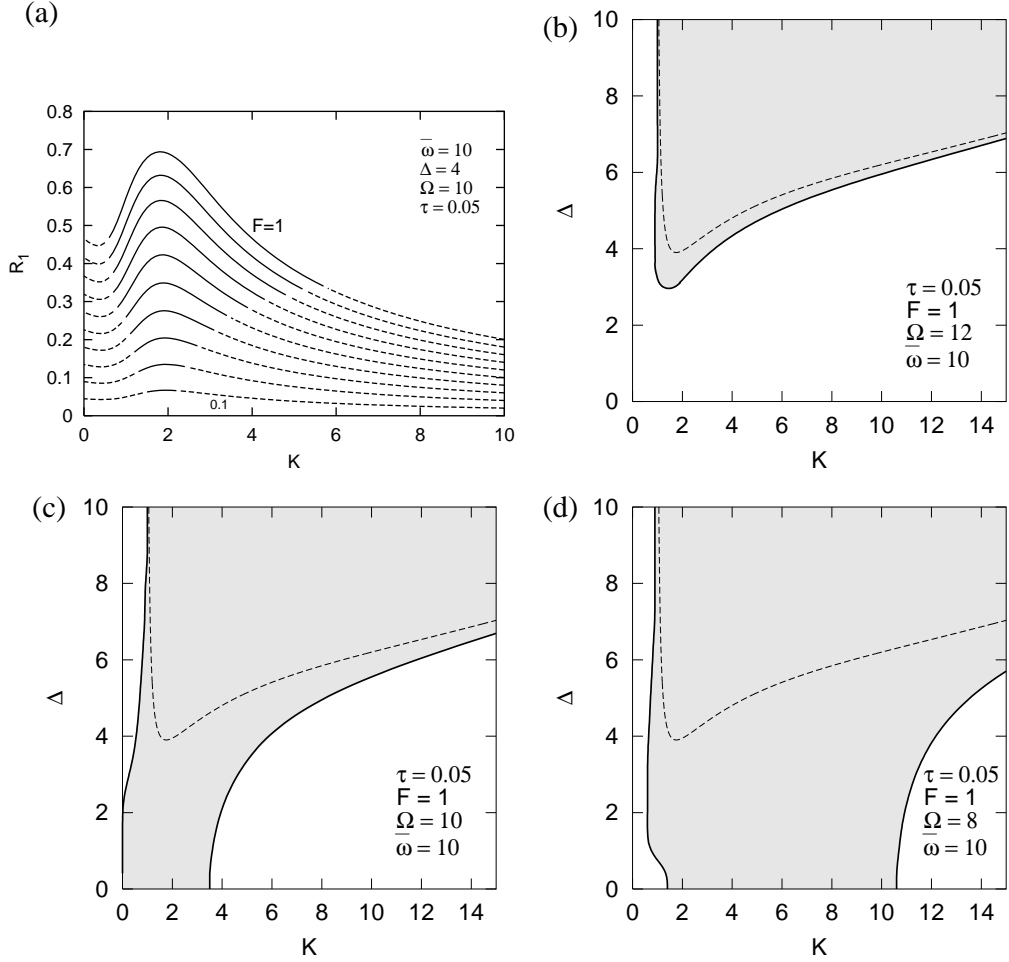}}}
\caption{\label{FIG:resp_KR_forF}%
(a) Response of the first oscillator for various $F$
as a function of $K$.
Stability regions (shaded) of the tuned state in $(K,\Delta)$ space
for (b) $\Omega=12$, (c) $\Omega=10$, (d) $\Omega=8$, 
while keeping the other parameters at
$F = 1$, $\tau = 0.05$, and $\bar{\omega} = 10$.}
\end{figure}

\begin{figure} 
\centerline{\scalebox{0.6}{\includegraphics{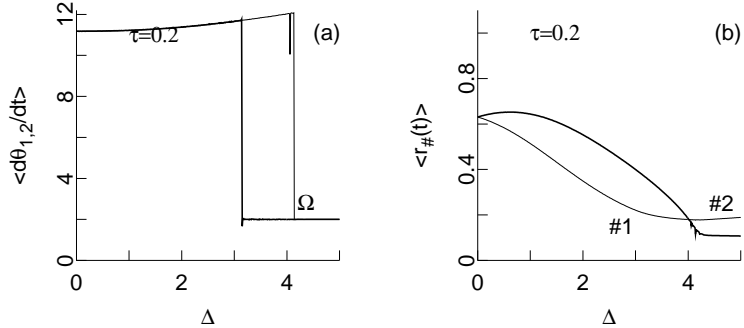}}}
\caption{\label{FIG:ave_delta}%
Average frequencies and the average amplitudes 
as a function of $\Delta$. The first of the two oscillators
has smaller frequency. In (a) the left vertical curve corresponds to 
the first oscillator.
$F = 1$, $\tau = 0.2$, $\bar{\omega} = 10$, $\Omega=2$, and $K=1.5$.}
\end{figure}

However, the amplitudes of the oscillators are in general not
identical. The response of the oscillators in the synchronized state is
unique only when $0\le\Omega\tau<2\pi$.
In Fig.~\ref{FIG:respvstau} the amplitudes and phases are shown
as a function of $\tau$. For a given $\tau$, stronger coupling
among oscillators decreases the amplitudes. At $\Omega \tau = \pi$,
the oscillators are locked with the external driver with
a maximum phase difference.
For a given set of parameters the external forcing
will widen the region of stability of the synchronized solution.
In Fig.~\ref{FIG:resp_KR_forF}(a) the response of one of the oscillators
is plotted for different values of $F$.
The boundary of the stability regions
of the death state in the absence of external driving
falls inside the stability region of the synchronized state.
In Figs.~\ref{FIG:resp_KR_forF}(b), (c), and (d),
the stability region
of the synchronized state is shown in $(K,\Delta)$ space for
three different values of the driving frequency $\Omega$ and
a fixed set of $F = 1.0$, $\tau = 0.05$, and $\bar{\omega} = 10$.
In each of these figures, the dashed curve indicates the marginal
stability curve for $F=0$.
As is evident the stability region of the tuned state is sensitive to
$\Omega$ and appears to increase in area for slow driving.
The difference between the effects of force (as seen from
Fig.~\ref{FIG:CritKdeltanodelay}) 
and that of time delay is evident here in the bifurcation diagram.
Time delay breaks up the degenerate bifurcation point
where as just the presence of force alone would not remove the degeneracy.
Finally we examine the average frequencies as a function
of the frequency dispersion. In the absence of 
time delay, as we noted in the previous section,
there is a transition of the average frequencies 
to the external frequency as $\Delta$ is increased.
However, in the presence of time delay, the intrinsic
frequencies are affected by $\tau$ differently, and thus
there could be connected regions of $\tau$ where the oscillators
can be found in two different locked states as shown in 
Fig.~\ref{FIG:ave_delta}. Here the first oscillator with
small frequency (left curve in the frequency plot)
is more susceptible to the external driving and gets locked
to it at a smaller value of $\Delta$ while the second
one is still locked to the branch of the average intrinsic 
frequency.  The corresponding amplitudes are shown in 
Fig.~\ref{FIG:ave_delta}(b). This feature is absent for
$\tau=0$ and may be found in multiple connected regions of 
$\tau$.

%
%
\subsection{Tuning property of time delay}
Having now found the region of stability of the tuned solutions,
we must then examine the response of the oscillators in this region
and find where the maximum or resonant response occurs and with
what functional relation this response scales with the driving force, $F$.
The Hopf bifurcation curves of the non-driven system in the
presence of finite $\tau$ have a special significance in that
on them the response is nonlinear even for small $F$. This fact was
noted earlier in the literature and was used to model
the nonlinear compression of sound frequencies by
the inner hair cells of the cochlea \cite{CDJP:00,EOCHM:00}.
In the presence of $\tau$ the frequency on such critical
curves has a dependence on $\tau$. This makes it possible to choose
a proper time delay such that a nonlinear response at a given
external frequency is achieved.

\begin{figure} 
\centerline{\scalebox{0.8}{\includegraphics{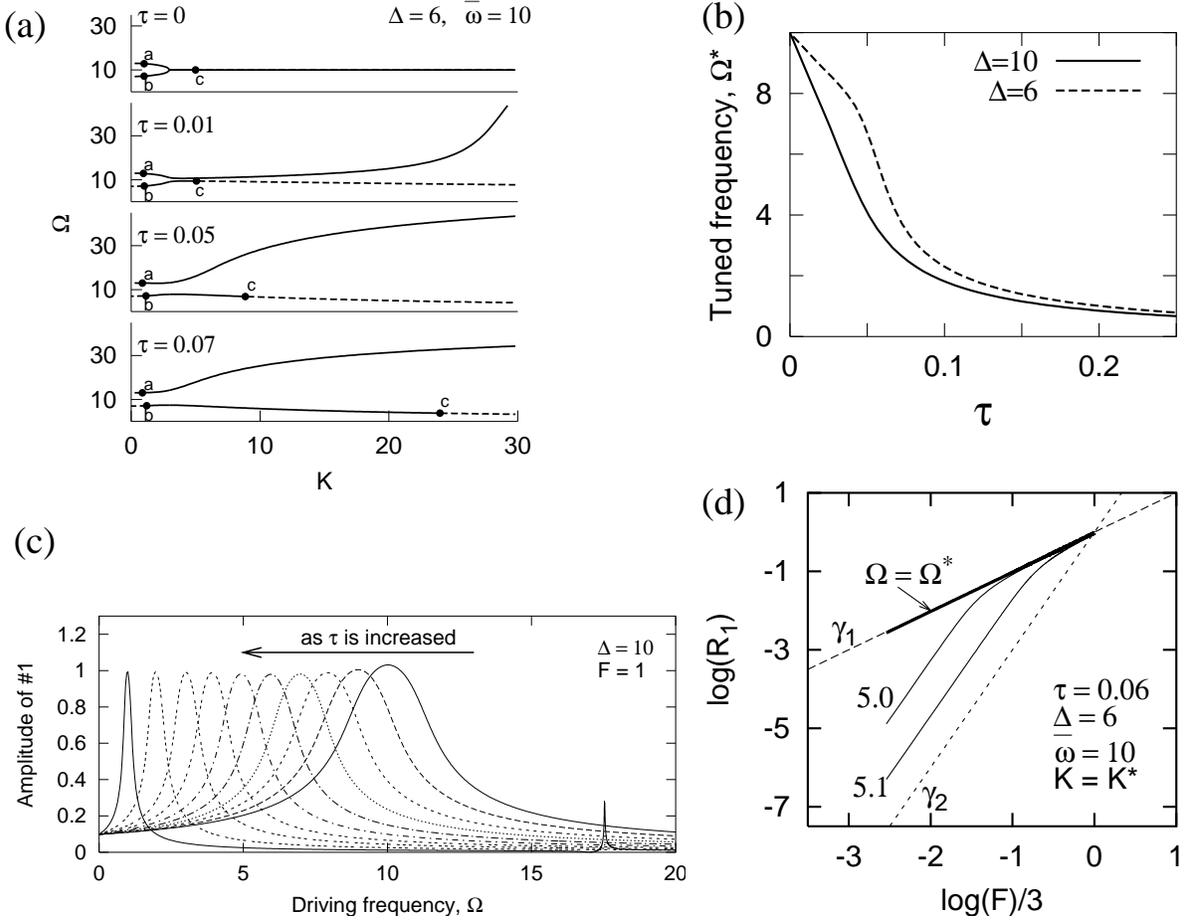}}}
\caption{\label{FIG:freq_spec2}%
(a) Two of the intrinsic frequencies existing in the undriven system.
The solid portions of the curves represent damped regions,
i.e. the real part of the corresponding eigenvalue has negative
sign. The bullets indicate the values of $K$ at which the
damping of that particular mode is zero,
(b) The tuned frequency as a function of $\tau$.
(c) Resonances on the tuned frequency curve.
(d) The amplitude response of one of the oscillators on and around the
tuned frequency curve. The curve $\gamma_1$ is a straight line with
slope unity, and $\gamma_2$ is that with 3.}
\end{figure}

We explore this interesting feature further by
examining the eigenvalue spectrum
of the death state in the absence of external force.
The eigenvalue equation that determines the stability of the origin is
\be
\label{EQN:eval2undrivenev}
\lambda = 1 - K \pm \sqrt{K^2{\mathrm e}^{-2\lambda \tau} -
\Delta^2/4} \pm \bar{\omega}
\ee
which can be derived from Eq.~\ref{EQN:eval2} by inserting
$R_{1,2} = 0$. We are now interested in the imaginary parts of
the eigenvalues. The real parts only reveal the damping of that
particular frequency mode when the driving is turned on.
A detailed study of the marginal stability was earlier carried
out in literature.
Here we examine the imaginary parts of the eigenvalue spectrum (i.e. the
intrinsic frequencies of the system)
as a function of $K$. As can be verified from the full spectrum,
the lower frequencies are least damped and the higher frequencies are
heavily damped. So we focus our attention on the first two frequencies
of the system. Since the eigenvalues occur in complex conjugate pairs,
it is sufficient to plot the positive frequencies.
In Fig.~\ref{FIG:freq_spec2}(a), the intrinsic
frequencies of the eigenvalue spectrum are plotted
while showing the nature of the damping existing
for each branch. The system responds linearly to the external driving
in the region of damping.
The two frequencies plotted have critical points on them marked
$a$, $b$ and $c$.
A nonlinear growth of the amplitude of the oscillators
occurs at all the points $a$, $b$, and $c$ as a function of the driving
force at any given value of $\Omega$.
The higher frequency branch has only one critical point $a$
at which the damping is zero. To the left of the point, the
frequency is a growing mode and to the right, the frequency is
a damped mode. Also note that the region left to point $a$ indicates
a drift region. The lower branch has two critical points on it: $b$ and $c$
both representing zero growth of the mode. The region left to the point
$b$ and that to the right of $c$ are growing modes. The critical points
$a$ and $b$ are not sensitive to variations in $\tau$ and thus their
frequency span is too limited. However the point $c$ is sensitive to $\tau$
and in fact spans a range of $\Omega$ values as $\tau$ is increased.
Thus this point is of interest to us.  At higher values of $\Delta$,
a large frequency range can be spanned with a short variation in $\tau$.
In Fig.~\ref{FIG:freq_spec2}(b) the contour of $c$ is plotted
in $(\Omega, \tau)$ space. At this frequency the system responds
nonlinearly and can be termed as the tuned frequency.
In Fig.~\ref{FIG:freq_spec2}(c) the frequency response of the
system is shown as one traverses along the tuned frequency
curve shown in Fig.~\ref{FIG:freq_spec2}(b) for $\Delta = 10$.

Now we show the response of the oscillators at the tuned
frequencies. The system responds nonlinearly with the
amplitude proportional to $F^{1/3}$. In Fig.~\ref{FIG:freq_spec2}(d)
the amplitude of one of the oscillators is plotted
in log-scale for a choice of the parameters lying on the
tuned frequency curve shown in Fig.~\ref{FIG:freq_spec2}(b):
$K (= K^*) = 16.861411237$, $\tau = 0.06$,
$\Omega = (\Omega^*) = 4.968163$, $\bar{\omega} = 10$,
$\Delta = 6$. The responses around the curve are linear, but
on the critical curve the response is nonlinear and is
proportional to $F^{1/3}$.

%
%
\section{Synchronization of a large number of coupled oscillators}
\label{SEC:large}
We now briefly consider the question of the possibility of 
synchronizing a large group of globally coupled oscillators
that have a distribution of frequencies, $g(\omega)$,
and are driven by an external periodic force: 
\bea
\dot{Z}_j(t) = (1+i\omega_j-\left|Z_j(t)\right|^2)Z_j(t) 
+ \frac{K}{N}\sum\limits_{k=1}^{N} [ Z_k(t-\tau)-Z_j(t)] 
+ F {\mathrm e}^{i\Omega t} ,
\eea
The contribution of the self coupling term that would emanate from
the summation above would be ignorable as the size of the system becomes 
large. 
A full bifurcation 
picture of these equations is relegated to a future study.
Here we present the main features exhibited by this system
which are common to the case of two oscillators discussed so far.
$g(\omega)$ is assumed to be in the form of uniform density distribution, i.e.
$g(\omega) = 1/2\gamma$ for $\omega\in[-\gamma,\gamma]$, and $0$
otherwise. The frequencies are chosen with equal spacing. Since
time delay removes the symmetry enjoyed by the system, a constant
positive value $\bar\omega$ is added to the frequency distribution
to make all the frequencies positive.
A convenient macroscopic parameter to characterize the collective properties of 
a large population of oscillators
is the order parameter defined by
\be
S = R(t) {\mathrm e}^{i\Phi(t)}= \frac{1}{N}\sum\limits_{j=1}^N Z_j(t) .
\ee
\begin{figure} 
\centerline{\rotatebox{0}{\scalebox{0.8}{\includegraphics{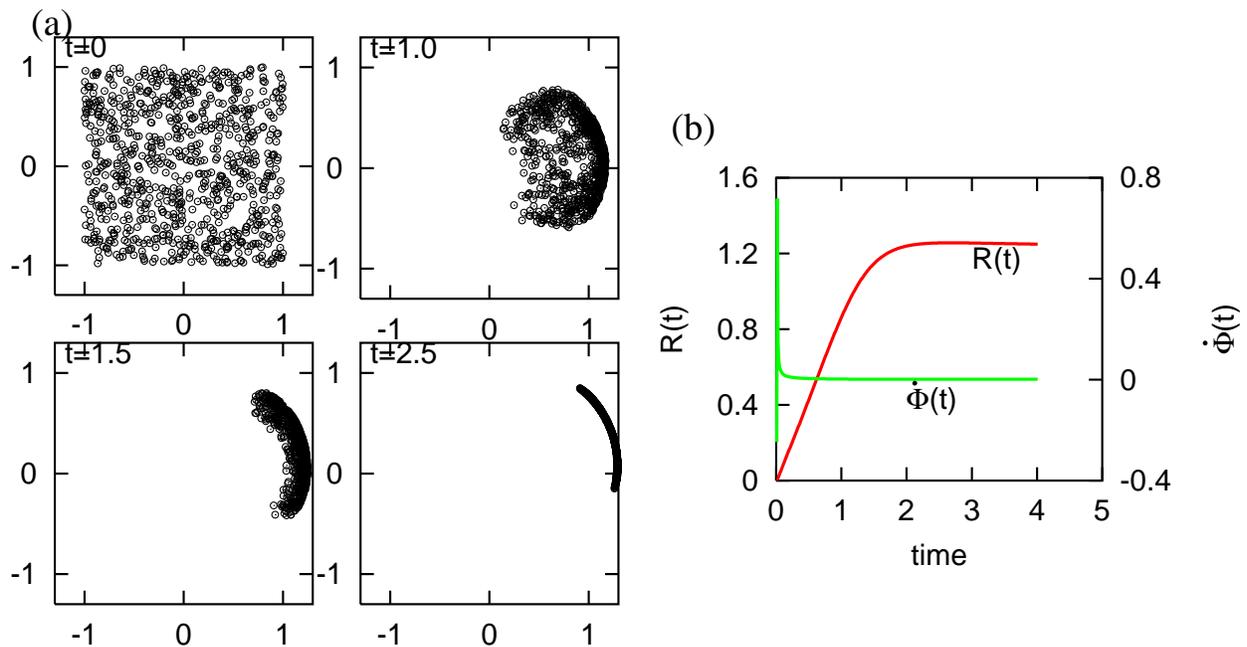}}}}
\caption{\label{FIG:N800club}%
(a)$(X,Y)$ snapshots of time evolution of 
$800$ coupled oscillators under external force.
(b) The evolution of the amplitude and the derivative of the phase
of the order parameter in the moving frame.
$\gamma=0.6$, $K=0.8$, $F=1$, $\tau=0.18$, $\Omega=1.0$, and $\bar{\omega}=1.2$.
}
\end{figure}
In Fig.~\ref{FIG:N800club}(a) the time evolution of $N=800$ oscillators
are plotted in $(X,Y)$ plane in the frame moving with the external frequency
for $F=1$, $\tau=0.18$, $\bar\omega=1.2$, and $\Omega=1.0$.
A stationary asymptotic state in the fourth panel indicates a complete
synchronization of all the oscillators in the population to the driving
frequency. Since the forcing is strong, the synchronization time
is short, and the coupling strength is overcome by $F$. 
In Fig.~\ref{FIG:N800club}(b) magnitude ($R$) and
the frequency ($\dot\Phi$ of the order parameter
are plotted in the moving frame. This confirms the synchrony.

\begin{figure} 
\centerline{\rotatebox{0}{\scalebox{0.8}{\includegraphics{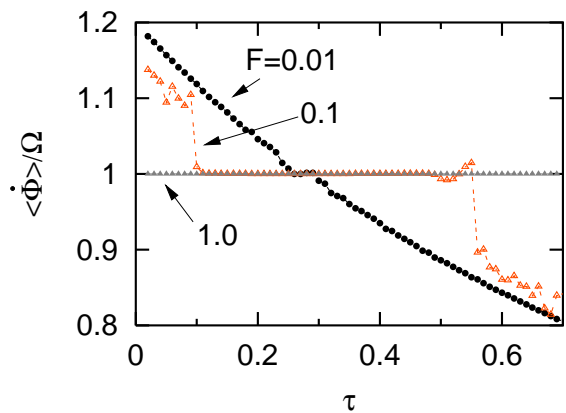}}}}
\caption{\label{FIG:N100tau}%
Synchronization regions for different $F$ as a function of $\tau$.
$N=100$, $\gamma=0.6$, $K=0.8$, $\Omega=1.0$, and $\bar{\omega}=1.2$.}
\end{figure}
One of our important results in the earlier sections is the
possibility of synchronization for infinitesimal forces in the presence 
of time delay. For this we carry out simulations on $N=100$
for finite time delay, and finite frequency spread for various values
of driving force. The results are presented in Fig.~\ref{FIG:N100tau}
as a plot of $<\dot{\Phi}>/\Omega$ vs. $\tau$. For a strong force ($F=1$)
the horizontal line indicates synchronization. The synchronization occurred
here even for $\tau=0$ and extends to a very large value of $\tau$.
As we decrease the force, synchronization at $\tau=0$ is lost, but
is attained for a range of $\tau$ values. Even for $F=0.01$, this particular
choice of $\gamma$ shows a range of $\tau$ values where synchronization is
found. This confirms our earlier results on two coupled oscillators.
Further work on the bifurcation structure of the large number of 
oscillators will be reported elsewhere.

\begin{figure} 
\centerline{\rotatebox{0}{\scalebox{1.0}{\includegraphics{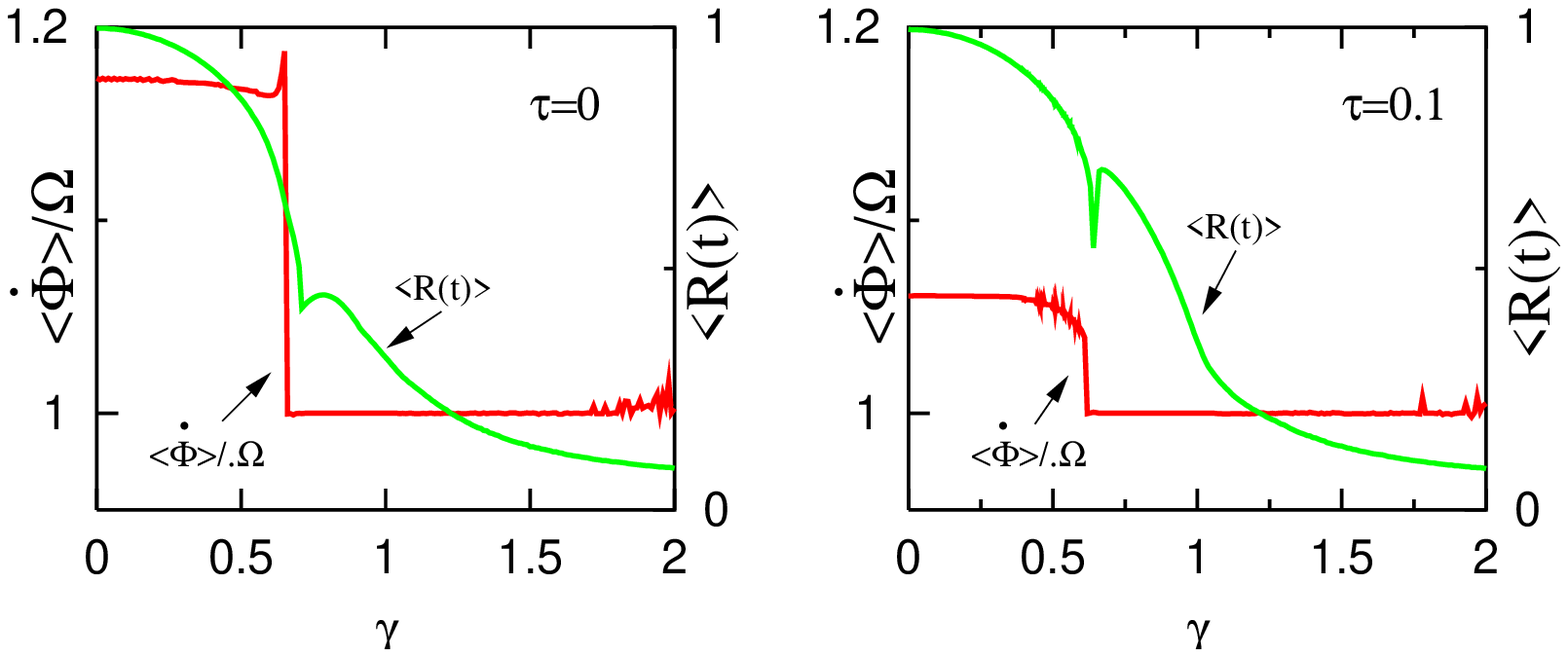}}}}
\caption{\label{FIG:opN100}%
Frequency jumps shown as a function of $\gamma$ for (a) $\tau=0$, and 
(b) $\tau=0.1$.
The other parameters are 
$N=100$, $K=0.8$, $F=0.1$, $\Omega=1$, and $\bar\omega=1.2$.
}
\end{figure}

Finally we look at the frequency transitions from synchronization
between the oscillators to that to the driving frequency.
In Fig.~\ref{FIG:opN100}(a) and (b) the average values of 
the amplitude of the order parameter
$<R(t)>$, and the ratio of the average frequency to the driving frequency,
$<\dot\Phi>/\Omega$ are plotted as a function of the frequency width
$\gamma$ for $\tau=0$, and $\tau=0.1$. 
At lower values of $\gamma$, as is also evident even for two 
coupled oscillators, the oscillators are not synchronized to the driver,
but make a transition to the synchronized state at a critical value.
At large values of $\gamma$ synchronization loses its stability.
The frequency of the collective state at lower $\gamma$ is suppressed
by time delay.

%
%
%
%
\section{Conclusions}
\label{SEC:con}
We have carried out a detailed theoretical analysis of the driven
periodic response of a system of two time delay coupled limit
cycle oscillators that are subjected to identical external
forces. In particular, we have examined the stability properties
of periodic phase locked oscillations of the system that are
synchronized with the external driving frequency. Bifurcation
diagrams in the parameter space of the coupling strength and
frequency spread of the two oscillators have been presented and
their variation with time delay has been discussed. We have also
obtained the characteristics of the response function in various
limits and highlighted their sensitivity to time delay. We have further 
confirmed that some of the important results of the two oscillator system 
also emerge in simulations carried out on a large number
of globally coupled limit cycle oscillators.

For the sake of completeness and to gain an overall perspective
of our results we have also showed the connection of our work to
earlier investigations of non-driven coupled oscillators, with
\cite{RSJ:00} or without \cite{AEK:90} time delay, and to driven
oscillators without time delay \cite{GH:83book}. The parametric
region corresponding to the {\it amplitude death} of two coupled
non-driven oscillators has been the primary area of our interest. This
is the region that sustains a periodic oscillation when the
system is driven externally. For our time delayed system the
extent of this region and the nature of
the driven response within it are all sensitive functions of the
driving frequency, driving amplitude and the time delay
parameter. This is evident from the detailed results presented in
the various previous sections. In general, the size of the
stability region of the forced response is larger than the size of
the original {\it death} region of the non-driven system. However
the bounding curves of the original region play a special role in
that the small amplitude response on them is highly nonlinear.\\

Two important results of our work are worth highlighting in view of 
their potential applications. First, by an appropriate choice of time delay,
it is possible to mitigate the disadvantages of far from resonance driving
and to achieve synchronization with a minimal of force.
Second, for a fixed driving frequency, the driven response of the system can be tuned
to any desired frequency with a suitable choice of the delay parameter.
This property relies on the nonlinear response of the system on the
Hopf bifurcation boundaries. Since the location of these boundaries 
(in the frequency space) is a sensitive function of time delay it provides 
for this possibility of using the time delay parameter as a control knob for frequency
tuning. It would be interesting to look for evidences of these effects operating
in natural biological and chemical systems or to implement them in appropriately
designed electronic receiver systems.\\

Finally we would like to remark that in our analysis 
we have not examined the driven response in the region of phase
drift solutions. Such a study is presently in progress as is the
extension to a larger system of coupled oscillators where regions
of chaotic solutions could be explored. Future work should also 
focus on large number of locally coupled oscillators.

%
%
\appendix*
\section{Derivation of Curves $\Gamma_1$ to $\Gamma_6$}

In order to obtain the bifurcation diagram in $(F,\tilde{\omega})$
plane, we must find the critical curves of the above equations
by setting Re$(\lambda) = 0$.
For the sake of convenience consider two cases:
(i) ${R}^2 > \left|\tilde{\omega}\right|$, and
(ii) ${R}^2 \le \left|\tilde{\omega}\right|$.
In either of these two cases the real part of the eigenvalues of
Eq.~(\ref{EQN:wwFFeval1}) 
are always less by a positive value $2K$
than those of the
Eq.~(\ref{EQN:wwFFeval2}).
Thus, at the critical set of parameters where the first two
equations have their real parts zero, the real parts of the other
two equations are negative.
Hence, it is obvious that
the boundaries of the steady state are completely determined by
the eigenvalues of Eq.~(\ref{EQN:wwFFeval1})
alone. In fact these two equations
determine the stability of the curve $\gamma$.
The other two equations provide critical curves that represent transitions
of the eigenvalues. We find that the other solutions the system admits
will merge with the in-phase solutions at these critical lines.
First we derive the critical curves under both the cases mentioned
above for
Eq.~(\ref{EQN:wwFFeval1}).
Then we treat the other two eigenvalue equations.

Considering Eq.~(\ref{EQN:wwFFeval1}) with $+$ sign 
under case (i),
${\mathrm Re}(\lambda) = 1-2{R}^2 + \sqrt{{R}^4-\tilde{\omega}^2}$, and
${\mathrm Im}(\lambda) = 0$.
The criticality occurs at
${\mathrm Re}(\lambda) = 0$ (and thus $1-2{R^*}^2 < 0$),
which results in
$R^2 = {R^*}^2_\pm = 2/3 \pm \sqrt{1/9-\tilde{\omega}^2/3}$.
Now by making use of Eq.~(\ref{EQN:Req2}), and the condition
$1-2{R^*_-}^2 < 0$, the following critical curve ($\Gamma_1$)
can easily be written down:
\be
  \label{EQN:gamma1}
  \Gamma_1 :  ~~~~F = \left\{\frac{2}{27}\left[(1+9\tilde{\omega}^2)+
   (1-3\tilde{\omega}^2)^{3/2}\right]\right\}^{1/2},
    ~~~\tilde{\omega}^2 \in [\frac{1}{4},\frac{1}{3}].
\ee
Similarly, making use of Eq.~(\ref{EQN:Req2}), and the condition
$1-2{R^*_+}^2 < 0$, the following critical curve ($\Gamma_2$)
can be written down:
\be
  \label{EQN:gamma2}
  \Gamma_2 :  ~~~~F = \left\{\frac{2}{27}\left[(1+9\tilde{\omega}^2)-(1-3\tilde{\omega}^2)^{3/2}\right]\right\}^{1/2},
   ~~~\tilde{\omega}^2 \in [0,\frac{1}{3}].
\ee
Under the case (i) again, Eq.~(\ref{EQN:wwFFeval1}) with $-$ sign yields
${\mathrm Re}(\lambda) = 1-2{R}^2 - \sqrt{{R}^4-\tilde{\omega}^2}$, and
${\mathrm Im}(\lambda) = 0$.
The criticality occurs at
${\mathrm Re}(\lambda) = 0$ (and thus $1-2{R^*}^2 > 0$),
which again results in
$R^2 = {R^*}^2_\pm = 2/3 \pm \sqrt{1/9-\tilde{\omega}^2/3}$.
Note that $R^2 = {R^*_+}^2$ fails to obey the condition
$1-2{R}^2 > 0$. Hence $R^2 = {R^*_{+}}^2$ does not result in
any critical curves. But $R^2={R^*_-}^2$ results in the following
critical curve:
\be
  \label{EQN:gamma1a}
  \Gamma_1 :  ~~~~F = \left\{\frac{2}{27}\left[(1+9\tilde{\omega}^2)
  +(1-3\tilde{\omega}^2)^{3/2}\right]\right\}^{1/2}, ~~~\tilde{\omega}^2 \in [0,\frac{1}{4}].
\ee
Under case (ii) 
Eq.~(\ref{EQN:wwFFeval1}) with both signs
can be considered together:
${\mathrm Re}(\lambda) = 1-2 R^2$, and
${\mathrm Im}(\lambda) = \pm \sqrt{\tilde{\omega}^2-R^4}$.
The criticality occurs at $R^2 = {R^*}^2 = 1/2$ with the
condition that $R^2<\left|\tilde{\omega}\right|$. This, when
substituted in Eq.~(\ref{EQN:Req2}),
results in the following critical curve:
\be
\label{EQN:gamma3}
\Gamma_3: ~~~~F = \left\{(\tilde{\omega}^2+1/4)/2\right\}^{1/2},
~~~\tilde{\omega}^2 \in [\frac{1}{4},\infty).
\ee
Since the imaginary part of the eigenvalue spectrum is nonzero,
the nature of the bifurcation across this critical curve is determined
by
\be
\frac{d}{dF}{\mathrm Re}\lambda \Big|_{{\mathrm Re}\lambda=0}
= \frac{-4 F}{\tilde{\omega}^2 - \frac{1}{4}}
\cases{> 0  & ~~if~~  $\tilde{\omega}^2 < \frac{1}{4}$, \cr
       < 0  & ~~if~~  $\tilde{\omega}^2 > \frac{1}{4}$,}
\ee
which indicates that the bifurcation is of the supercritical
Hopf kind.

We now consider
Eq.~(\ref{EQN:wwFFeval2}) with $+$ sign
under case (i):
${\mathrm Re}(\lambda) = 1 - 2K - 2R^2 + \sqrt{R^4-\tilde{\omega}^2}$, and
${\mathrm Im}(\lambda) = 0 $.
Let us define $f(\rho) = \{\rho^3-2 \rho^2+(1+\tilde{\omega}^2)\rho\}^{1/2}$,
and $a=2K-1$.
Following exactly the same method as described above,
the critical curves can be derived as follows for $K < \frac{1}{2}$:
\bea
\label{EQN:gamma4}
\Gamma_4: F = \{f(\rho_-) ~\Big|~
\rho_- = \frac{-2 a}{3} - \sqrt{\frac{a^2}{9}-\frac{\tilde{\omega}^2}{3}}\},
~~\tilde{\omega}^2 \in[a^2/4,a^2/3],\\
\label{EQN:gamma5}
\Gamma_5: F = \{f(\rho_+) ~\Big|~
\rho_+ = \frac{-2 a}{3} + \sqrt{\frac{a^2}{9}-\frac{\tilde{\omega}^2}{3}}\},
~~\tilde{\omega}^2 \in[0,a^2/3].
\eea
And using the Eq.~(\ref{EQN:wwFFeval1}) with $-$ sign, the corresponding curve
turns out to be:
\bea
\label{EQN:gamma4a}
\Gamma_4: F = \{f(\rho_-) ~\Big|~
\rho_- = \frac{-2 a}{3} - \sqrt{\frac{a^2}{9}-\frac{\tilde{\omega}^2}{3}}\},
~~\tilde{\omega}^2 \in[0,a^2/4].
\eea
Still under case (i), the nature of the curves under $K> \frac{1}{2}$
differ from the above. The curves are derived using the same method as
above.
Eq.~(\ref{EQN:wwFFeval1}) with $-$ sign results in
\bea
\label{EQN:gammac4}
\Gamma_4: F = \{f(\rho_-) ~\Big|~
\rho_- = \frac{-2 a}{3} - \sqrt{\frac{a^2}{9}-\frac{\tilde{\omega}^2}{3}}\},
~~\tilde{\omega}^2 \in[0,a^2/3]. \\
\label{EQN:gammac5}
\Gamma_5: F = \{f(\rho_+) ~\Big|~
\rho_+ = \frac{-2 a}{3} + \sqrt{\frac{a^2}{9}-\frac{\tilde{\omega}^2}{3}}\},
~~\tilde{\omega}^2 \in[a^2/4,a^2/3],
\eea
and the Eq.~(\ref{EQN:wwFFeval1}) with $+$ sign results in
\be
\label{EQN:gammac5a}
\Gamma_5: F = \{f(\rho_+) ~\Big|~
\rho_+ = \frac{-2 a}{3} + \sqrt{\frac{a^2}{9}-\frac{\tilde{\omega}^2}{3}}\},
~~\tilde{\omega}^2 \in[0,a^2/4].
\ee
Now we are left with case (ii). The equation
(\ref{EQN:wwFFeval1}) with both the signs 
can be treated together:
${\mathrm Re}(\lambda) = 1 - 2K - 2R^2 $ and
${\mathrm Im}(\lambda) = \pm \sqrt{R^4-\tilde{\omega}^2}$.
On the critical curve, $R^2 = {R^*}^2 = \frac{1}{2}-K$.
This gives rise to the following critical curve in $(F,\tilde{\omega})$ plane:
\be
\label{EQN:gamma6}
\Gamma_6: F = \Big\{(\frac{1}{2}-K) [ \tilde{\omega}^2 + (K+\frac{1}{2})^2\Big\}^{1/2}, ~~~
K < \frac{1}{2}, ~~\tilde{\omega}^2 \in [\frac{1}{2}-K, \infty).
\ee
Since the imaginary part of the eigenvalue is nonzero in this case,
the nature of transition of eigenvalues is determined by
\be
\frac{d{\mathrm Re}\lambda}{dF} \Big|_{{\mathrm Re}\lambda = 0}
= \frac{-4F}{ \tilde{\omega}^2 - (\frac{1}{4}-K-3 K^2) }
\cases{
> 0 & ~~if~~ $ \tilde{\omega}^2 < \frac{1}{4}-K-3 K^2$, \cr
< 0 & ~~if~~ $ \tilde{\omega}^2 > \frac{1}{4}-K-3 K^2$,}
\ee
which indicates that the bifurcation involved is once again a supercritical Hopf
bifurcation.

%
%

\end{document}